\documentclass[twocolumn,
 superscriptaddress,
 amsmath,amssymb,
 aps,
 pra,
 floatfix,
]{revtex4-2}

\usepackage[T1]{fontenc}
\usepackage[utf8]{inputenc}

\usepackage{dsfont}

\usepackage{physics}
\usepackage{circuitikz}
\usepackage{graphicx}

\usepackage[breaklinks=true,colorlinks,citecolor=blue,linkcolor=blue,urlcolor=blue]{hyperref}

\usepackage[capitalise]{cleveref}

\newcommand{\sz}{\hat \sigma_z}
\newcommand{\sy}{\hat \sigma_y}
\newcommand{\sx}{\hat \sigma_x}

\newcommand{\aop}{\hat a}
\newcommand{\adop}{\hat a ^\dagger}
\newcommand{\bop}{\hat b}
\newcommand{\bdop}{\hat b ^\dagger}

\newcommand{\figref}[1]{\mbox{Fig.~\ref{#1}}}
\newcommand{\figpanel}[2]{Fig.~\hyperref[#1]{\ref*{#1}(#2)}}
\newcommand{\figurepanel}[2]{Figure~\hyperref[#1]{\ref*{#1}(#2)}}
\newcommand{\figpanels}[3]{Figs.~\hyperref[#1]{\ref*{#1}(#2)-(#3)}}

\begin{document}
\title{Renormalization and Low-Energy Effective Models in Cavity and Circuit QED}

\author{Daniele Lamberto}
\thanks{These authors contributed equally.}
\affiliation{Dipartimento di Scienze Matematiche e Informatiche, Scienze Fisiche e  Scienze della Terra, Universit\`{a} di Messina, I-98166 Messina, Italy}

\author{Alberto Mercurio}
\thanks{These authors contributed equally.}
\affiliation{Laboratory of Theoretical Physics of Nanosystems (LTPN), Institute of Physics, Ecole Polytechnique Fédérale de Lausanne (EPFL), CH-1015 Lausanne, Switzerland}
\affiliation{Center for Quantum Science and Engineering, EPFL, CH-1015 Lausanne, Switzerland}

\author{Omar Di Stefano}
\affiliation{Dipartimento di Scienze Matematiche e Informatiche, Scienze Fisiche e  Scienze della Terra, Universit\`{a} di Messina, I-98166 Messina, Italy}

\author{Vincenzo Savona}
\affiliation{Laboratory of Theoretical Physics of Nanosystems (LTPN), Institute of Physics, Ecole Polytechnique Fédérale de Lausanne (EPFL), CH-1015 Lausanne, Switzerland}
\affiliation{Center for Quantum Science and Engineering, EPFL, CH-1015 Lausanne, Switzerland}

\author{Salvatore Savasta}
\affiliation{Dipartimento di Scienze Matematiche e Informatiche, Scienze Fisiche e  Scienze della Terra,
	Universit\`{a} di Messina, I-98166 Messina, Italy}

\date{\today}

\begin{abstract}
    The quantum Rabi model (QRM) is a cornerstone in the study of light-matter interactions within cavity and circuit quantum electrodynamics (QED). It effectively captures the dynamics of a two-level system coupled to a single-mode resonator, serving as a foundation for understanding quantum optical phenomena in a great variety of systems.
    However, this model may produce inaccurate results for large coupling strengths, even in systems with high anharmonicity. Moreover, issues of gauge invariance further undermine its reliability. In this work, we introduce a renormalized QRM that incorporates the effective influence of higher atomic energy levels, providing a significantly more accurate representation of the system while still maintaining a two-level description.
    To demonstrate the versatility of this approach, we present two different examples: an atom in a double-well potential and a superconducting artificial atom (fluxonium qubit). This procedure opens new possibilities for precisely engineering and understanding cavity and circuit QED systems, which are highly sought-after, especially for quantum information processing.
\end{abstract}

\maketitle

Effective low-energy models serve as foundational tools across diverse areas of physics, from describing many-body systems to characterizing light-matter interactions. By systematically eliminating terms that virtually excite particles into high-energy states through perturbative methods, they provide simplified yet accurate descriptions of low-energy physics while incorporating the influence of high-energy states through renormalized coupling terms. This approach has proven to be remarkably successful in various contexts. In condensed matter physics, notable examples include the effective spin Hamiltonian derived from the half-filled Hubbard model~\cite{Takahashi1977Half-filled,MacDonald1988tUexpansion} and the Kondo model~\cite{Schrieffer1966Relation}. The BCS theory of superconductivity \cite{BCS1957} and the theory of Josephson tunneling \cite{josephson1962possible} also rely heavily on effective low-energy descriptions. Beyond condensed matter, these techniques form the backbone of renormalization group methods in high-energy physics~\cite{Wilson1983}.

In the context of light-matter interactions, low-energy effective Hamiltonians serve as indispensable tools, particularly in cavity and circuit quantum electrodynamics (QED)~\cite{brune1996quantum,haroche2013nobellecture,Blais2021Circuit,DeBernardis2024Tutorial}. These models are typically derived by truncating the matter Hilbert space to a few relevant states while considering only one or a few modes of the electromagnetic resonator. The quantum Rabi model (QRM) stands as the quintessential example, describing the interaction between a two-level quantum emitter and a single-mode resonator. The QRM has provided deep insights into the physics of cavity and circuit QED systems, and has also been experimentally realized across diverse platforms, including superconducting qubits~\cite{wallraff2004strong,gu2017microwave}, semiconductor quantum dots~\cite{Reithmaier2004Strong,Lodahl2015interfacting}, and trapped ions~\cite{Leibfried2003Quantum}.

Although these models effectively describe such systems, there are regimes where the two-level truncation fails to produce reliable results. In superconducting qubits like the transmon \cite{Koch2007Charge_insensitive}, fluxonium \cite{Manucharyan2009Fluxonium}, or flux qubits \cite{Lloyd1999fluxqubit}, truncating to a few energy levels is common but often inaccurate, especially when the anharmonicity of such systems is not sufficiently high. This limitation affects critical operations in quantum information~\cite{Nielsen2012Quantum} such as readout, gate fidelities, and error correction, where precise modeling is essential. For instance, it has been shown that the readout of the transmon may ionize the system, leading to significant measurement errors~\cite{Shillito2022Dynamics,Dumas2024Measurement_Induced}. Another factor that can compromise the accuracy of the two-level projection is the coupling strength. Recent experimental advances have demonstrated how cavity and circuit QED systems can reach non-perturbative regimes, such as the ultrastrong coupling (USC) and deepstrong coupling (DSC) regimes~\cite{forndiaz2019ultrastrong,frisk2019ultrastrong}. In these regimes, the standard QRM faces several fundamental issues, including the breakdown of gauge invariance~\cite{debernardis2018breakdown,stokes2019gauge,distefano2019resolution,taylor2020resolution,savasta2021gauge,dmytruk2021gauge,settineri2021gauge}, complications in the theory of photodetection~\cite{ridolfo2012photon,mercurio2022regimes}, and challenges in treating open quantum systems~\cite{beaudoin2011dissipation,settineri2018dissipation,mercurio2023pure}. In all these cases, the standard QRM may yield inaccurate description of the behaviour of these devices, even in the presence of very high anharmonicity, thus calling for a more advanced model.

In this work, we propose a different perspective to address these issues through a renormalization procedure of the QRM, which we name Renormalized QRM (RQRM). 
In this effort, we were inspired by another well-established class of effective Hamiltonians widely adopted in the quantum theory of polaritons in solids, i.e. the Hopfield model and its few-mode generalizations, which describe the interaction between photons and collective matter excitations~\cite{hopfield1958theory,grossosolid2000}. In particular, it has been demonstrated that accurately describing such experimental results across various systems and spectral ranges (see, e.g., \cite{Hopfield1965dielectric, Frohlich1971observation}) requires the introduction of an Hopfield model, where both the photonic resonance frequency and the light-matter coupling strength are renormalized by introducing a background dielectric function $\epsilon_\infty$, which effectively accounts for the influence of higher-energy excitations.
While renormalizing models of interacting bosons is  simple due to their harmonic nature, the renormalization of the QRM is not as straightforward and remains largely unexplored.

In this work we propose a systematic approach for developing RQRMs, which account for the influence of higher atomic energy levels. We first derive a general formulation of the RQRM applicable to both cavity and circuit QED systems. Extensive numerical simulations demonstrate that these RQRMs significantly improve accuracy across different coupling regimes and anharmonicities. We then explore the connections between our approach and other theoretical methods, in particular the resolvent technique. Additionally, we investigate two more aspects of the RQRM, i.e. its gauge invariance properties and its impact on physical observables.

\begin{figure*}[t]
    \centering
    \includegraphics[width=\linewidth]{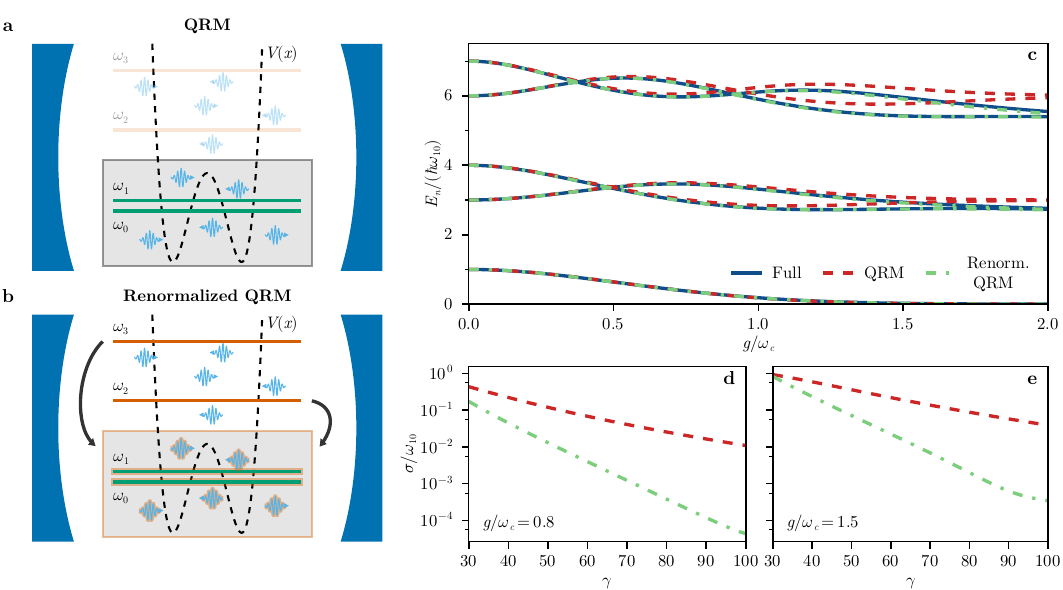}
    \caption{\textbf{Renormalization of the Quantum Rabi Model.} We consider a single electron in a one-dimensional double-well potential, interacting with a cavity mode. \textbf{a}, Standard procedure for obtaining the QRM from the full Hamiltonian, which consists of simply projecting in the low-energy subspace of the atomic eigenstates. \textbf{b}, Renormalization of the QRM, which takes into account the interaction of photons with the higher-energy atomic levels. The renormalization procedure leads to the presence of new terms in the effective low-energy Hamiltonian, providing more accurate results, still treating the system as a two-level system. \textbf{c}, Comparison of the eigenvalues in the full model (blue solid), the standard QRM (red dashed), and the RQRM (green dash-dotted). The RQRM provides more accurate results, even for strong coupling strengths. Parameters used: $m=1$, $\gamma=60$, and $\omega_c=3\omega_{10}$. \textbf{d}-\textbf{e}, Mean square error of the eigenvalues of the first 5 excited states with respect to the full model, as a function of the anharmonicity parameter $\gamma$, for $g / \omega_c = 0.8$ (\textbf{d}) and $g / \omega_c = 1.5$ (\textbf{e}). 
    }
    \label{fig: main-figure}
\end{figure*}

\vspace{0.5cm}
{\noindent \Large \textbf{Results}}
\\
{\textbf{Renormalization of the Quantum Rabi Model in Cavity QED}}

In the context of cavity QED, the QRM can be derived from a variety of different systems. As a pedagogical example, we start by considering a single electric dipole of charge $q$ and mass $m$ in a one-dimensional potential, which interacts with a single cavity mode with frequency $\omega_c$. 
In the electric-dipole approximation, the radiation wavelength is much larger than the atomic size, allowing us to neglect the spatial dependence of the vector potential, i.e. $\hat A = A_0 \left( \aop + \adop \right)$, where $A_0$ is the zero-point fluctuation amplitude and $\aop \left(\adop\right)$ is the annihilation (creation) operator of the electromagnetic mode. The Hamiltonian in the dipole gauge reads~\cite{power1959coulomb,woolley1971molecular,babiker1983derivation,cohen-tannoudji1997photons_lagrangian}
\begin{align}
    \label{eq:H_full_D}
    \hat H_{\rm D} =& \ \frac{\hat p^2}{2 m} + V(\hat x) + \hbar \omega_c \adop \aop \nonumber \\
    &- i q \omega_c A_0 \hat x \left( \aop - \adop \right) + \frac{\omega_c q^2 A_0^2}{\hbar} \hat x^2 \, .
\end{align}
where $V(\hat x)$ is the atomic potential, $\hat x$ is the position operator and $\hat p$ is its conjugate momentum. For illustrative purposes, let us assume the case of a double-well potential $V(\hat x) = \alpha \hat x^4 - \beta \hat x^2$, with $\alpha, \beta > 0$. 
By adjusting the parameter $\gamma = m \beta^3 / (\hbar^2 \alpha^2)$, we can tune the anharmonicity of the atomic system (see Supplementary Section 2).
In particular, increasing $\gamma$ leads to an increase in the ratios $\omega_{jk} / \omega_{10}$, where $\omega_{jk} = \omega_j - \omega_k$ is the transition frequency between the $j$-th and $k$-th atomic levels.

The standard QRM is obtained by truncating the atomic Hilbert space to the two lowest energy levels, which can be formally obtained by applying the projection operator $\hat P = \sum_{n=0}^1 \dyad{n}{n}$ to the full Hamiltonian (see \figref{fig: main-figure}(a)). It was shown that the QRM in the dipole gauge yields more accurate results compared to other possible gauges~\cite{debernardis2018breakdown}. Therefore, the standard QRM Hamiltonian in the dipole gauge reads
\begin{equation}  \label{eq:H_QRM}
    \hat{\mathcal{H}}_{\rm D} = \hat{P} \hat H_{\rm D} \hat{P}  = \frac{\hbar \bar{\omega}_{10}}{2} \sz + \hbar \omega_c \adop \aop - i \hbar g_{01} \sx \left( \aop - \adop \right) \, ,
\end{equation}
where $\bar{\omega}_{10} = \omega_{10} + (G_{11} - G_{00})/\omega_c$ is the two-level resonance frequency renormalized by the $\hat x^2$ contribution, as $G_{jk} = \sum_{l} g_{jl} g_{lk} \equiv \omega_c^2 q^2 A_0^2 \mel{j}{\hat x^2}{k} / \hbar^2$. The terms $g_{jk} = \omega_c q A_0 \mel{j}{\hat x}{k} / \hbar$ are the light-matter coupling strengths and $\hat \sigma_i$ are the Pauli matrices. It is worth mentioning that the projection operation can be performed in various ways, leading to different variations of $\bar{\omega}_{10}$~\cite{debernardis2018breakdown,distefano2019resolution,arwas2023metrics} (see Supplementary Section 1). 
The projection onto the unperturbed atomic states, as in \cref{eq:H_QRM}, was recently shown to give more accurate results~\cite{arwas2023metrics}.

The standard QRM in \cref{eq:H_QRM} is well-known to closely match the full model when $\bar{\omega}_{10} \simeq \omega_c$ or for weak coupling strengths~\cite{lamb1952fine,bassani1977choice}. However, when these conditions are not met, its accuracy progressively worsens~\cite{debernardis2018breakdown}.
 
To address these limitations, we now ask whether an improved version of the QRM can be developed, still within a two-level description, leading us to derive the RQRM. To this end, we first apply a Schrieffer-Wolff (SW) transformation~\cite{Schrieffer1966Relation,Bravyi2011schriefferwolff} to the full Hamiltonian in Eq.~\eqref{eq:H_full_D} (see Methods). 
The chosen transformation effectively treats the high-energy subspace as a perturbation, while the two-level subspace is handled non-perturbatively, marking a key distinction from traditional perturbative methods.
Indeed, this is possible because the higher-energy atomic levels are in the dispersive regime with the cavity, characterized by $|g_{jk} / (\omega_{jk} - \omega_c)| \ll 1$. This condition is satisfied for large enough atomic anharmonicities, which ensure that the transitions from the two lowest energy levels to the higher ones are not resonant with the cavity frequency, i.e. $\abs{\omega_{jk}} \gg \omega_c$.
This approach allows for a fully non-perturbative treatment of the QRM, where the two-level system and the cavity field are effectively dressed by the contributions of higher atomic levels (see \figref{fig: main-figure}(b)). 
The resulting Hamiltonian reads
\begin{align}
    \label{eq:H_QRM_renorm}
    \hat{\mathcal{H}}_\mathrm{D}^\mathrm{eff} = & \frac{\hbar \tilde{\omega}_{10}}{2} \sz + \hbar \omega_c \adop \aop \nonumber 
    + \hbar \left( B_+ + B_- \sz \right)\left( \aop - \adop \right)^2 \nonumber \\ 
    & - i \hbar \tilde{g}_{01} \sx \left( \aop - \adop \right) - D_{01} \sy \left( \aop + \adop \right) \, ,
\end{align}
with $\tilde{\omega}_{10} = \bar{\omega}_{10} + 2 A_-$ and $\tilde{g}_{01} = g_{01} + C_{01}$ being the renormalized two-level resonance frequency and coupling strength, respectively. Here, $A_\pm$, $B_\pm$, $C_{01}$ and $D_{01}$ are the renormalization parameters defined in \crefrange{eq:app-A_coeff}{eq:app-D_coeff} of the Methods Section, which depend on the microscopic details of the high energy states.
Notably, the RQRM in \cref{eq:H_QRM_renorm} provides a significantly more accurate description of the full system without requiring explicit enlargement of the Hilbert space.

Figure \ref{fig: main-figure}(c) shows a comparison of the eigenvalues of the first 5 excited states obtained from the full model $\hat{H}_\mathrm{D}$ (solid blue line), the standard QRM in \cref{eq:H_QRM} (red dashed), and the RQRM in \cref{eq:H_QRM_renorm} (green dash-dotted). The results confirm that the RQRM yields more accurate predictions than the standard QRM, which diverges at relatively high coupling strengths. The coupling strength $g = \abs{g_{01}}$ was varied through $A_0$, which proportionally scales all the other coupling strengths $g_{jk}$ as well.

In \cref{fig: main-figure}(d-e), we show the mean square error with respect to the full model $\sigma = \sqrt{\sum_{j=1}^N \left( E_j - E_j^\mathrm{full} \right)^2 / N}$ of the eigenvalues of the first $N=5$ excited states of the full light-matter system. We compare the QRM with the RQRM as a function of the anharmonicity parameter $\gamma$ for two different coupling strengths, namely $g / \omega_c = 0.8$ and $g / \omega_c = 1.5$. We observe that the RQRM not only provides more accurate results, but also scales better with increasing anharmonicity.

\begin{figure*}[t]
    \centering
    \includegraphics{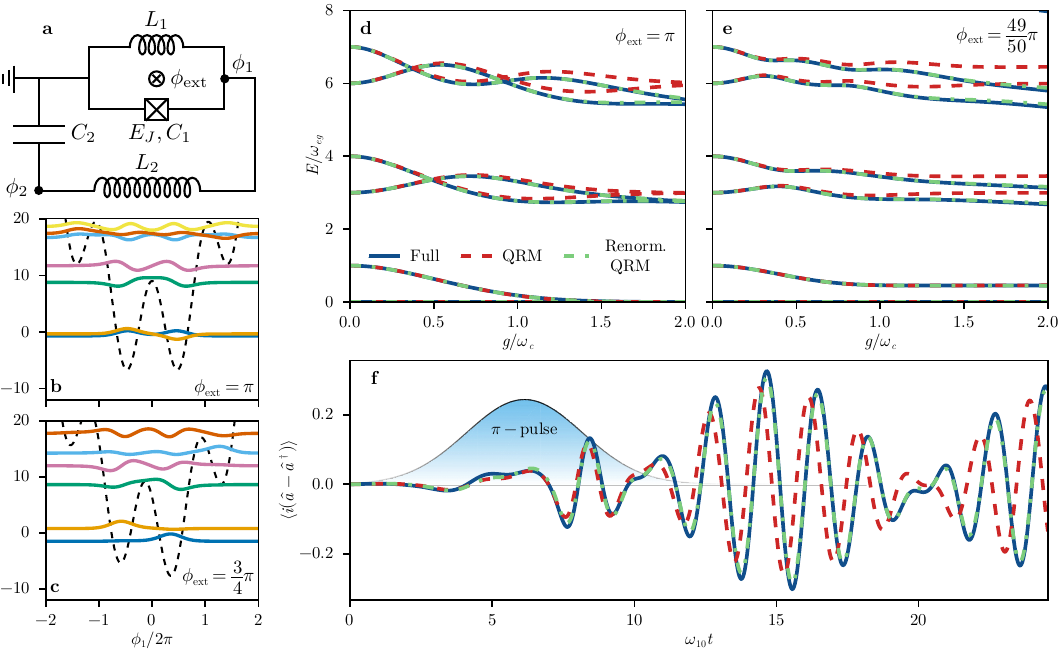}
    \caption{\textbf{Renormalization of the fluxonium circuit.} \textbf{a}, Schematical representation of a fluxonium qubit, defined by a Josephson junction with energy $E_J$ and a capacitance $C_1$ in parallel with a inductance $L_1$, galvanically coupled to a $LC$ resonator. \textbf{b-c}, First eigenstates of the fluxonium qubit for $\phi_\mathrm{ext}=\pi$ (\textbf{b}) and $\phi_\mathrm{ext}= 3 \pi / 4$ (\textbf{c}). In the latter, the parity symmetry is broken. In both cases, the dashed black line correspond to the potential. \textbf{d-e}, Comparison of the eigenvalues in the full model (solid blue line), the standard QRM (dashed green line), and the RQRM (dotted red line), as a function of the normalized coupling $g / \omega_c$, and for $\phi_\mathrm{ext}=\pi$ (\textbf{d}) and $\phi_\mathrm{ext}= 49 \pi / 50$ (\textbf{e}). As for the real atoms case, the renormalization gives better results. \textbf{f}, Time evolution of $\expval{i (\aop - \adop)}$ after a $\pi$-pulse on the qubit and in the case of $\phi_\mathrm{ext}=49\pi/50$. The renormalized QRM provides a better agreement with the full model. The parameters used in this Figure are: $E_C = q^2 / (2 C_1) = 2.5 \ \mathrm{GHz}$, $E_L = (\hbar / 2 q)^2 / L_1 = 0.5 \ \mathrm{GHz}$, $E_J = 9 \ \mathrm{GHz}$, and $\omega_c = 3 \omega_{10}$, which reproduce typical experimental values for fluxonium qubits~\cite{Manucharyan2009Fluxonium,Mencia2024Integer}. For the $\pi$-pulse, we used $\omega_\mathrm{dr} = E_{10}$, $\sigma_\mathrm{dr} = 50 / (E_{21} - E_{10})$ and $t_0 = 3 \sigma_\mathrm{dr}$ (see Methods).}
    \label{fig:circuits}
\end{figure*}

It is interesting to observe that some of the renormalization terms in \eqref{eq:H_QRM_renorm} become negligible for high anharmonicity, namely the $B_-$ and $D_{01}$ terms (see Methods). Therefore, we can write a simplified version of the RQRM, which is given by
\begin{align}
    \label{eq:H_QRM_D_renormalized_simplified}
    \hat{\mathcal{H}}_\mathrm{D}^{\prime} = & \frac{\hbar \tilde{\omega}_{10}}{2} \sz + \hbar \omega_c \adop \aop \nonumber \\
    & + \hbar B_+ \left( \aop - \adop \right)^2 - i \hbar \tilde{g}_{01} \sx \left( \aop - \adop \right) \, .
\end{align}

We can now perform a Bogolioubov transformation on the above Hamiltonian, which allows us to diagonalize the purely photonic terms. The resulting Hamiltonian reads
\begin{equation}
    \label{eq:H_QRM_renorm_bogoliubov}
    \hat{\mathcal{H}}_\mathrm{D}^\prime = \frac{\hbar \tilde{\omega}_{10}}{2} \sz + \hbar \tilde{\omega}_c \bdop \bop - i \hbar \tilde{g}_{01} \sqrt{\frac{\omega_c}{\tilde{\omega}_c}} \sx \left( \bop - \bdop \right) \, ,
\end{equation}
where $\tilde{\omega}_c = \sqrt{\omega_c^2 - 4 B_+ \omega_c}$ is the renormalized cavity frequency, and $\hat{b}-\hat{b}^\dagger = \sqrt{\tilde{\omega}_c / \omega_c} ( \aop - \adop )$. Remarkably, \cref{eq:H_QRM_renorm_bogoliubov} preserves a QRM-like structure with renormalized frequencies and coupling, although in general providing slightly less accurate results, with  respect to \cref{eq:H_QRM_renorm}.
This result explains why the QRM still remains a valid description even in the deepstrong coupling regime, when the main parameters are fitted from the experimental data and not derived from a microscopic theory~\cite{yoshihara2017superconducting}.

\vspace{0.5cm}
{\noindent \textbf{Renormalization of the Quantum Rabi Model in Circuit QED}}

Superconducting circuits are a promising platform for quantum technologies, as they allow the realization of ultrastrong and even deepstrong coupling between artificial atoms and electromagnetic modes~\cite{niemczyk2010circuit,yoshihara2017superconducting,Krantz2019A,Blais2021Circuit}. In this context, to illustrate our approach, we focus on a specific example involving the fluxonium qubit.
In particular, we focus our attention on the study of a fluxonium-resonator system~\cite{Manucharyan2009Fluxonium,Manucharyan2017Resilience,Nguyen2019High_Coherence} represented in \figref{fig:circuits}(\textbf{a}), bearing in mind that the same procedure can be applied to other circuits, once the corresponding Hamiltonian is derived. The fluxonium qubit is characterized by higher anharmonicity when compared to other devices, such as the transmon qubit, making it an ideal candidate for the study of the renormalization of the QRM. 
Specifically, the system under study is composed by a Josephson junction with energy $E_J$ and capacitance $C_1$ in parallel with a inductance $L_1$ to which is added in series an $L_2 C_2$ resonator. An external flux $\phi_{\rm ext}$ is threading the loop formed by the Josephson junction and the inductance $L_1$.
The corresponding Hamiltonian is derived by the usual quantization procedure~\cite{Vool2017Introduction,Blais2021Circuit} (see Supplementary Section 3), resulting in
\begin{equation}
    \label{eq:H_circuit}
    \hat{H} = \hat{H}_{\rm flux} \left( \hat{\phi}_1 , \hat{Q}_1 \right) + \hat{H}_{\rm res} \left( \hat{\phi}_2 - \hat{\phi}_1 , \hat{Q}_2 \right) \, ,
\end{equation}
where $\hat{H}_{\rm flux}$ and $\hat{H}_{\rm res}$ are the fluxonium and resonator bare Hamiltonians, given by
\begin{alignat}{2}
    \label{eq:H_flux}
    \hat{H}_{\rm flux} \left( \hat{\phi}_1 , \hat{Q}_1 \right) & = \frac{\hat{Q}_1^2}{2 C_1}  + \frac{\hat{\phi}_1^2}{2 L_1} - E_J \cos{\left( \frac{\hat{\phi}_1 - \phi_{\rm ext}}{\phi_0} \right)} \, , \\
    \label{eq:H_res}
    \hat{H}_{\rm res} \left( \hat{\phi}_2 , \hat{Q}_2 \right) & = \frac{\hat{Q}_2^2}{2 C_2} + \frac{\hat{\phi}_2^2}{2 L_2} \, ,
\end{alignat}
respectively, with $\phi_0 = \hbar / 2e$ being the reduced flux quantum and $\phi_i$ ($Q_i$) the flux (charge) node variables. 
Equation \eqref{eq:H_res} can be rewritten as $\hat{H}_{\rm res} = \hbar \omega_c \adop \aop$, with bosonic operator $\aop = (\omega_c \hat{\phi}_2 + i \hat{Q}_2) / \omega_c \phi_{\rm zpf}$, $\omega_c = 1/\sqrt{L_2 C_2}$ and $\phi_\mathrm{zpf} = \sqrt{\hbar / 2 C_2 \omega_c}$.

The Hamiltonian in \cref{eq:H_circuit} exhibits a striking similarity to the cavity-QED Hamiltonian in \cref{eq:H_full_D}. Indeed, it can be shown that \cref{eq:H_full_D} can be obtained by performing a minimal coupling replacement on the photonic operators defined by $\aop \to \aop + i q A_0 \hat{x} / \hbar$ \cite{garziano2020gauge,settineri2021gauge}. Analogously, \cref{eq:H_circuit} presents a minimal coupling replacement on the resonator variables. The primary distinction between the two cases lies in the nature of the coupling. In the former, the interaction term is of the form coordinate-momentum $(\hat{x}\hat{E})$, while in the latter, it involves the two fluxes, which correspond to the generalized coordinates in the so-called flux gauge. Therefore, the two Hamiltonians are simply linked by a unitary transformation $\hat T = \exp \left( i \pi \adop \aop / 2 \right)$ (see Methods). Consequently, the same procedure used in the previous section can be applied here.

First, we derive the Hamiltonian for the QRM by projecting the full Hamiltonian in the low-energy subspace of the fluxonium qubit, which results in
\begin{align}
    \label{eq:H_fluxres_QRM}
    \hat{\mathcal{H}}_{\rm fr} = & \ \hbar \omega_c \adop \aop + \hbar \bar{\omega}_{10} \sz + \hbar \frac{G_{01}}{\omega_c} \sx \nonumber \\
    & - \hbar \left( \frac{g_{11} + g_{00}}{2} \hat{I} + \frac{g_{11} - g_{00}}{2} \sz + g_{01} \sx \right) \!\! \left( \aop + \adop \right)
\end{align}
where
\begin{equation} \label{eq:coupling_circuit}
    g_{jk} = \frac{\phi_\mathrm{zpf} \phi_{jk}}{\hbar L_2} \, , \quad
    G_{jk} = \frac{\hbar L_2 \omega_c}{2 \phi_\mathrm{zpf}^2} \sum_l g_{jl} g_{lk} = \frac{\omega_c}{2 \hbar L_2} \Phi_{jk} \, ,
\end{equation}
with $\phi_{ij} = \bra{i}\hat{\phi}_1\ket{j}$ and $\Phi_{ij} = \bra{i}\hat{\phi}_1^2\ket{j} = \sum_k \phi_{ik}\phi_{kj}$.
Alongside the renormalized qubit frequency $\bar{\omega}_{10}$ (previously defined), three additional terms emerge in \cref{eq:H_fluxres_QRM} due to the breaking of parity symmetry, originating from the external flux $\phi_{\rm ext}$. These additional terms vanish when $\phi_\mathrm{ext}=k \pi$ (with $k \in \mathbb{Z}$), where the symmetry is restored. The third term arises from the projection of $\hat{\phi}_1^2$ onto the unperturbed fluxonium states, similarly to what discussed in the cavity QED case.

Following our approach, we can apply the renormalization procedure to the fluxonium-resonator system. After performing the SW transformation (see Methods), the renormalized Hamiltonian is given by
\begin{align} \label{eq:H_fluxres_renorm}
    \hat{\mathcal{H}}_{\rm fr}^{\rm eff} = & \ \hbar \omega_c \adop \aop + \hbar \tilde{\omega}_{10} \sz + \hbar \left( \frac{G_{01}}{\omega_c} + A_{10} + A_{01} \right) \sx \nonumber \\
    - & \hbar \left( \frac{\tilde{g}_{11} + \tilde{g}_{00}}{2} \hat{I} + \frac{\tilde{g}_{11} - \tilde{g}_{00}}{2} \sz + \tilde{g}_{01} \sx \right) \left( \aop + \adop \right) \nonumber \\
    - & \hbar \left( B_+ \hat{I} + B_- \sz + 2 B_{01} \sx \right) \left( \aop + \adop \right)^2 \nonumber \\
    + & i \hbar \left(A_{10} - A_{01}\right) \sy \left( \aop^2 - \hat{a}^{\dagger^2} \right) - i \hbar D_{01} \sy \left( \aop - \adop \right) \, ,
\end{align}
with $\tilde{\omega}_{10} = \bar{\omega}_{10} + 2 A_-$ and $\tilde{g}_{jk} = g_{jk}+ C_{jk}$. It is worth noting that all the coefficients are the same as those in the natural atom case \cref{eq:H_QRM_renorm}, defined explicitly in the Methods Equations (\ref{eq:app-A_coeff}-\ref{eq:app-D_coeff}). The Hamiltonian in \cref{eq:H_fluxres_renorm} is the Circuit QED version of the RQRM.

Figure~\ref{fig:circuits}(\textbf{b-c}) show the shapes of the fluxonium potential and the respective eigenstates for the symmetric (\textbf{b}) and asymmetric (\textbf{c}) cases, corresponding to $\phi_\mathrm{ext} = \pi$ and $\phi_\mathrm{ext} = 3 \pi / 4$, respectively. \figref{fig:circuits}(\textbf{d-e}) show the comparison of the eigenvalues of the first 5 excited states obtained from the full model (blue solid), the standard QRM (red dashed), and the RQRM (green dash-dotted), as a function of the normalized coupling $g / \omega_c$, and for both the symmetric (\textbf{d}) and asymmetric (\textbf{e}) cases. 
Figure \eqref{fig:circuits}(\textbf{f}) shows the time evolution of the expectation value $\langle i (\hat{a} - \hat{a}^\dagger) \rangle$, considering the system initially in its ground state and applying a $\pi$-pulse  on the qubit, in presence of symmetry breaking (see Methods). 
These results demonstrate that the RQRM provides significantly more accurate results than the standard QRM, even in absence of parity symmetry in the anharmonic potential [\cref{fig:circuits}(\textbf{e})]. Furthermore, the RQRM performs better also in describing the system dynamics and its impact on the observables. In particular, this aspect will be further examined in the following sections.

\vspace{0.5cm}
\noindent \textbf{Higher-order corrections of the Effective Hamiltonians}

The effective Hamiltonians both in cavity [\cref{eq:H_QRM_renorm}] and circuit QED [\cref{eq:H_fluxres_renorm}] are obtained by performing a SW transformation up to the second order in the $g_{jk} / (\omega_{jk} - \omega_c)$ expansion. This leads to an improvement in the accuracy, as can be seen from \cref{fig: main-figure}(\textbf{c-e}) and \cref{fig:circuits}(\textbf{d-f}). One may wonder whether higher-order corrections can further improve the accuracy of the renormalization procedure. Notice that, even expanding the series of the SW to higher orders, the original eigenvalues are not recovered because the transformation $\hat{H}\rightarrow e^{-\hat{S}} \hat{H} e^{\hat{S}}$ preserves the spectrum of the Hamiltonian only if no projection is performed.

In order to quantify the impact of higher-order corrections on the renormalization, here we adopt the resolvent method~\cite{CohenTannoudji1998AtomPhoton,Mila2010Strong-Coupling,Fulde2012electron_book}.
In particular, we focus on the cavity QED model, but the same analysis can be applied to other cases. Starting from the eigenvalue equation of the full Hamiltonian in the dipole gauge $\hat{H}_\mathrm{D} \ket{\psi_\mathrm{D}} = E \ket{\psi_\mathrm{D}}$, it is possible to derive an eigenvalue equation for an effective Hamiltonian, defined in the projected subspace, as
\begin{equation}
    \label{eq:resolvent-eigenvalue-equation}
    \hat{\mathcal{H}}_\mathrm{D}^{(\mathrm{res.})} (E) \ket{\psi_\mathrm{D}} = E \ket{\psi_\mathrm{D}} \, ,
\end{equation}
where $E$ is the same eigenvalue as in the full case. The energy-dependent effective Hamiltonian is given by
\begin{equation}
    \label{eq:resolvent-effective-Hamiltonian}
    \hat{\mathcal{H}}_\mathrm{D}^{(\mathrm{res.})} (E) = \hat{P} \hat{H}_\mathrm{D} \hat{P} + \hat{P} \hat{H}_\mathrm{D} \hat{Q} \frac{1}{E - \hat{Q} \hat{H}_\mathrm{D} \hat{Q}} \hat{Q} \hat{H}_\mathrm{D} \hat{P} \, ,
\end{equation}
where $\hat{Q} = \hat{I} - \hat{P}$ is the projector on the subspace complementary to $\hat{P}$.

\begin{figure}
    \centering
    \includegraphics[width=\linewidth]{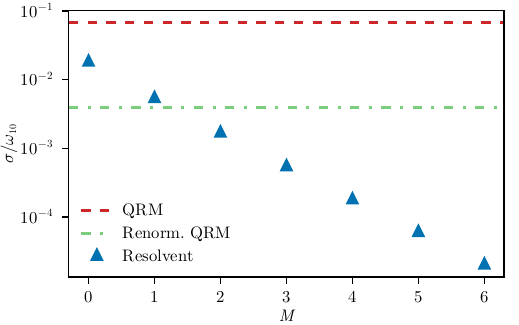}
    \caption{\textbf{Convergence of the Resolvent Method.} Mean square error of the first 5 excited states with respect to the full model, as a function of the order of the resolvent expansion. The error decreases exponentially with the order of the series $M$, showing the convergence of the resolvent method. To make a comparison, we show also the QRM and RQRM errors. We used the same parameters as in \cref{fig: main-figure}(c) with $g / \omega_c = 0.8$.}
    \label{fig: resolvent-convergence}
\end{figure}

Equation \eqref{eq:resolvent-eigenvalue-equation} gives the exact eigenvalues in the low-energy subspace. However, the implicit dependence of the effective Hamiltonian on the eigenvalue $E$ may make the problem challenging. 
Nonetheless, the resolvent $\hat{G} (E) \equiv (E - \hat{Q} \hat{H}_\mathrm{D} \hat{Q})^{-1}$ can be expanded in series by using the property $(A - B)^{-1} = A^{-1} \sum_{n=0}^\infty (B A^{-1})^n$, with  $A = \hbar \omega_0 - \hat{Q} \hat{H}_{0^\prime} \hat{Q}$ and $B = \hat{H}_\mathrm{int, D} - E + \hbar \omega_0$, where $\hat{H}_{0^\prime}$ and $\hat{H}_\mathrm{int, D}$ are defined in \cref{eq:app-H0_Schrieffer_Wolff} and \cref{eq: app - H_int_H}, respectively. The series can be truncated up to the $M$-\emph{th} order, obtaining a polynomial eigenvalue problem, which can be solved numerically~\cite{Mehrmann2004Nonlinear}. In the limit of $M \to \infty$, we recover the exact solution.

Figure \ref{fig: resolvent-convergence} shows the mean square error of the first 5 excited states with respect to the full model, as a function of the order of the resolvent expansion. We observe that the error decreases exponentially with $M$, showing the convergence of the method. As a comparison, we plot also the QRM and RQRM errors. While the resolvent method provides higher accuracy, it remains purely numerical. In contrast, the SW approach developed here enables the analytical derivation of an effective Hamiltonian.

\vspace{0.5cm}
\noindent \textbf{Gauge invariance of the Renormalized Quantum Rabi Model}

Here, we investigate the gauge properties of the RQRM. 
Models that do not preserve gauge invariance may lead to wrong predictions. One of the most striking examples is the Dicke model, where a collection two-level systems coupled to light was predicted to undergo a second-order phase transition to a photon condensate~\cite{dicke1954coherence,hepp1973superradiant,wang1973phase,emary2003quantum,emary2003chaos,buzek2005instability}. However, it was later shown that this phase transition is actually forbidden if the so-called $\hat{\mathbf{A}}^2$ term, which restores the gauge invariance, is not neglected~\cite{rzazewski2006comment,nataf2010no}. 
More generally, the photon condensation has been shown to be forbidden by the gauge invariance itself, both in truncated systems~\cite{andolina2022nogo} and in the full Hilbert space~\cite{knight1978are,andolina2019cavity,andolina2020theory}.
In the following, we will derive the RQRM in a gauge invariant form, and we will discuss the implications of this result.

\begin{figure}[b]
    \centering
    \includegraphics[width=\linewidth]{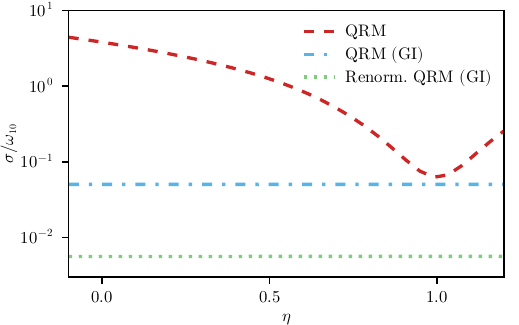}
    \caption{\textbf{Gauge invariance of the Renormalized QRM.} Mean square error of the eigenvalues of the first 5 excited states with respect to the full model, as a function of the gauge parameter $\eta$, for $g / \omega_c = 0.8$, $m=1$, $\gamma=60$, and $\omega_c=3\omega_{10}$. The QRM (red dashed) breaks gauge invariance, showing that the dipole gauge ($\eta = 1$) is the most accurate. On the other hand, the gauge invariant RQRM (green dotted) is not only gauge invariant but also provides more accurate results. For completeness, we also compare these models with the gauge-preserving QRM $\hat{\mathcal{H}}_\mathrm{D}$ (blue dash-dotted) derived in Ref.~\cite{savasta2021gauge}, which, however, does not take into account the renormalization of the higher energy levels.}
    \label{fig: gauge_comparison}
\end{figure}

As it is well-known~\cite{stokes2019gauge,arwas2023metrics}, in the full Hilbert space, we can introduce a gauge parameter $\eta$ which gives rise to a continuous family of $\eta$-dependent Hamiltonians $\hat H^{(\eta)}$ interpolating between the Coulomb and the dipole gauges, spanned by the unitary transformation $\hat U^{(\eta)} = \exp ( i \eta q \hat x \hat A / \hbar )$.
Explicitly, for a spatially constant vector potential, we have
\begin{equation} \label{eq:H_eta_dependent}
    \hat{H}^{(\eta)} = \hat{U}^{(1-\eta)} \hat{H}_\mathrm{a} \hat{U}^{(1-\eta) \dagger} + \hat{U}^{(\eta)} \hat{H}_\mathrm{ph} \hat{U}^{(\eta)} \, ,
\end{equation}
where $\hat{H}_\mathrm{a} = \hat{p}^2 / (2 m) + V (\hat{x})$ and $\hat{H}_\mathrm{ph} = \hbar \omega_c \adop \aop$ are the bare atomic and photonic Hamiltonians. For $\eta=0$, we recover the usual minimal coupling replacement in the Coulomb gauge Hamiltonian $(\hat{H}_\mathrm{C}=\hat{H}^{(0)})$, whereas for $\eta=1$ we obtain the dipole gauge Hamiltonian $(\hat{H}_\mathrm{D}=\hat{H}^{(1)})$, showing that in this case the minimal coupling is applied to the photonic term (notice that $\hat{U}^{(0)} = \hat{I}$).

When truncating to the atomic low-energy subspace, gauge invariance breaks down, as $\hat{\mathcal{H}}_\mathrm{C} = \hat{P} \hat{H}_\mathrm{C} \hat{P}$ and $\hat{\mathcal{H}}_\mathrm{D} = \hat{P} \hat{H}_\mathrm{D} \hat{P}$ are no longer linked by any unitary transformation.
This issue was resolved by applying the minimal coupling replacement directly in the projected Hilbert space~\cite{distefano2019resolution,taylor2020resolution,dmytruk2021gauge}. 
Indeed, analogously to the case of the full models, a unitary transformation in the truncated Hilbert space $\hat{\mathcal{U}} = \exp [ i q \hat{P} \hat{x} \hat{P} \hat{A}] = \exp [ i g_{01} \sx (\aop + \adop) / \omega_c ]$ is defined, which interpolates between the Coulomb and dipole gauges as 
\begin{equation}
    \hat{\mathcal{H}}^{(\eta)} = \hat{\mathcal{U}}^{(1-\eta)} \hat{\mathcal{H}}_\mathrm{a} \hat{\mathcal{U}}^{(1-\eta) \dagger} + \hat{\mathcal{U}}^{(\eta) \dagger} \hat{\mathcal{H}}_\mathrm{ph} \hat{\mathcal{U}}^{(\eta)} \, ,
\end{equation}
where $\hat{\mathcal{H}}_\mathrm{a} = \hbar \omega_{10} \sz / 2$.
This procedure restores a discrete form of gauge invariance directly within the truncated space \cite{savasta2021gauge}.

Following the same reasoning, the RQRM in \cref{eq:H_QRM_renorm_bogoliubov} can be interpreted in a gauge-preserving form as
\begin{equation}
    \label{eq:H_QRM_renorm_bogoliubov_GI}
    \hat{\mathcal{H}}_\mathrm{D}^\prime = \hat{\mathcal{H}}_\mathrm{a}^\prime + \hat{\mathcal{U}}^{\prime (1) \dagger} \hat{\mathcal{H}}_\mathrm{ph}^\prime \hat{\mathcal{U}}^{\prime (1)} \, ,
\end{equation}
where the renormalized atomic and photonic Hamiltonians are
\begin{equation} \label{eq:H_atom_ph_renormalized_GI}
    \hat{\mathcal{H}}_\mathrm{a}^\prime = \frac{\hbar \tilde{\omega}_{10}}{2} \sz \, , \quad\,\,
    \hat{\mathcal{H}}_\mathrm{ph}^\prime = \hbar \tilde{\omega}_c \bdop \bop \, ,
\end{equation}
and the renormalized $\eta$-dependent unitary operator is
\begin{equation}
    \label{eq:U_renormalized_GI}
    \hat{\mathcal{U}}^{\prime (\eta)} = \exp \left[ i \eta \frac{\tilde{g}_{01}}{\tilde{\omega}_c} \sqrt{\frac{\omega_c}{\tilde{\omega}_c}} \sx \left( \bop + \bdop \right) \right] \, .
\end{equation}
Consequently, the $\eta$-interpolated RQRM becomes
\begin{equation}
    \label{eq:H_QRM_renorm_GI}
    \hat{\mathcal{H}}^{\prime (\eta)} = \hat{\mathcal{U}}^{\prime  (1-\eta)} \hat{\mathcal{H}}_\mathrm{a}^\prime \hat{\mathcal{U}}^{\prime (1-\eta) \dagger} + \hat{\mathcal{U}}^{\prime (\eta) \dagger} \hat{\mathcal{H}}_\mathrm{ph}^\prime \hat{\mathcal{U}}^{\prime (\eta)} \, .
\end{equation}

Figure \ref{fig: gauge_comparison} shows the mean square error on the energies of the first five excited states, compared to those obtained from the full model as a function of the gauge parameter $\eta$, for $g / \omega_c = 0.8$. While the standard QRM breaks gauge invariance (with dipole gauge being the most accurate), the QRM $\hat{\mathcal{H}}^{(\eta)}$ and RQRM $\hat{\mathcal{H}}^{\prime (\eta)}$ ensure gauge invariance, with the latter providing a better accuracy.

\vspace{0.5cm}
{\noindent \textbf{The effect of the renormalization on the observables and matrix elements}}

In the previous sections, we have established that the renormalization significantly improves the energy spectrum of the system. However, a comprehensive understanding requires also examining how renormalization influences other physical quantities. In the following, we analyze how the renormalization affects both photonic and atomic observables and matrix elements in the cavity QED case.

\begin{figure}[t]
    \centering
    \includegraphics[width=\linewidth]{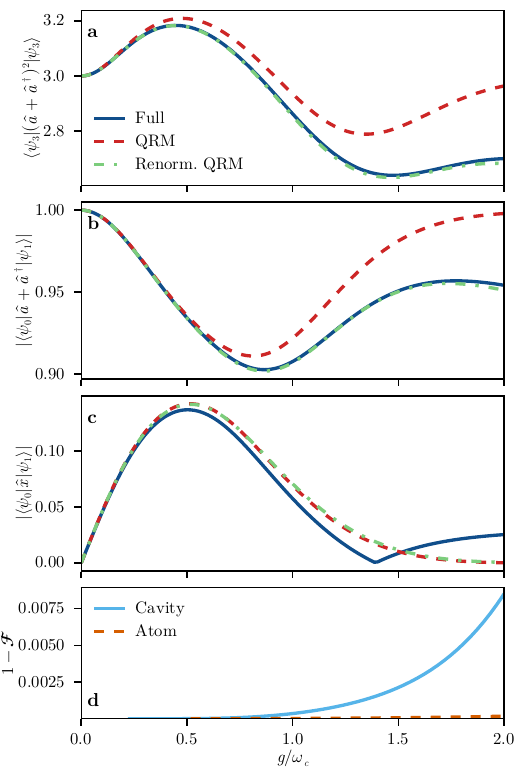}
    \caption{\textbf{Observables of the Renormalized QRM.} \textbf{a}, Expectation value of $(\aop + \adop)^2$ (\textbf{a}) on the third excited state of the full model (blue solid), QRM (red dashed) and the RQRM (green dash-dotted), as a function of the coupling strength $g / \omega_c$. The renormalized QRM provides more accurate results, even for strong coupling strengths. \textbf{b-c}, Matrix elements of the cavity field $\aop + \adop$ (\textbf{b}) and the atomic position operator $\hat{x}$ (\textbf{c}) between the ground and the second excited, as a function of $g / \omega_c$. While in the panel \textbf{b} the renormalization improves the accuracy, in the panel \textbf{c} it does not. This behavior can be explained by the infidelity of the second excited state of the RQRM with respect to the QRM, as a function of $g / \omega_c$, for both the reduced density matrix of the cavity (solid light blue) and the atom (dashed orange) (\textbf{d}).}
    \label{fig: observables}
\end{figure}

Figure \ref{fig: observables}(\textbf{a}) shows the expectation value $\bra{\psi_3}(\aop + \adop)^2\ket{\psi_3}$ on the third excited state, as a function of the coupling strength $g / \omega_c$. The RQRM provides more accurate predictions, even for large coupling strengths. A similar improvement is observed in \cref{fig: observables}(\textbf{b}), which displays the matrix element $|\mel{\psi_0}{(\aop + \adop)}{\psi_2}|$, between the ground and the second excited states. Although not directly observable, this quantity is closely related to dissipation rates and line broadening, which play an important role in experimental realizations~\cite{breuer2002theory,beaudoin2011dissipation,settineri2018dissipation}. 

Interestingly, \cref{fig: observables}(\textbf{c}) reveals that renormalization does not improve the accuracy of the matrix element $| \mel{\psi_0}{\hat{x}}{\psi_2} |$. 
This behavior can be in part understood by observing \cref{fig: observables}(\textbf{d}), which shows the infidelity $1-\mathcal{F}$ of $\ket{\psi_2}$ between the standard QRM and the RQRM, for the reduced density matrices of the cavity and atom, as a function of $g / \omega_c$. More specifically, the fidelity $\mathcal{F}$ is calculated as:
\begin{equation}
    \mathcal{F} = \Tr \left[ \sqrt{\sqrt{\hat{\rho}_{\mu}} \hat{\sigma}_{\mu} \sqrt{\hat{\rho}_{\mu}}} \right] \, ,
\end{equation}
where $\hat{\rho}_{\mu}$ and $\hat{\sigma}_{\mu}$ are the reduced density matrices for the renormalized and standard QRM, respectively, with $\mu$ indicating either the cavity or atom subsystems. The plot reveals that the renormalization does not significantly change the atomic subsystem, compared to the standard QRM, explaining the behavior in \figref{fig: observables}(\textbf{c}). On the contrary, the degrees of freedom of the cavity are strongly affected by the renormalization to more accurately reproduce the full model.
In fact, the operator $\hat{x}$ includes contributions from higher energy levels that cannot be reproduced by the corresponding operator projected into a two-level Hilbert space. Furthermore, the two-level truncation inherently discretizes space~\cite{savasta2021gauge}, introducing intrinsic inaccuracies in atomic observables that persist even after renormalization. These differences explain the contrasting behaviors observed in panels (\textbf{a-b}) and (\textbf{c}).

\vspace{0.5cm}
{\noindent \Large \textbf{Discussions}}

In this work, we have demonstrated how renormalizing the QRM by taking into account higher atomic energy levels leads to more accurate predictions, while maintaining analytical and computational tractability. This improvement is particularly evident in the energy spectrum and expectation values of observables, where the RQRM shows better agreement with the full model compared to the standard QRM. We have also derived a RQRM preserving gauge invariance, a fundamental physical requirement often violated by truncated models. As shown in the fluxonium-qubit example, renormalized low-energy effective Hamiltonians can also accurately capture symmetry-breaking effects while preserving the simplicity of a two-level description. This makes it especially useful for studying quantum information protocols~\cite{Nielsen2012Quantum}, particularly in the context of superconducting systems~\cite{niemczyk2010circuit,yoshihara2017superconducting,Krantz2019A,Blais2021Circuit}, where precise modeling of light-matter interactions is essential for predicting gate fidelities and decoherence effects.

We expect renormalization to have an even stronger impact on systems characterized by multi-dimensional potentials, such as flux-qubits~\cite{Lloyd1999fluxqubit}. Additionally, this procedure can be extended to other systems in cavity and circuit QED, e.g., systems with multiple non-linear components, or generalized models incorporating three, four, or more low-energy levels while still accounting for the influence of higher-energy states. Such extensions could be particularly valuable for systems where intermediate energy levels play a significant role, such as in Lambda or ladder-type atomic configurations commonly encountered in quantum optics \cite{You2011superconducting,Fleischhauer2005Lambda}.
Additionally, this renormalization approach could be extended to multi-mode cavity systems. This method could also be adapted for other hybrid quantum systems, e.g., optomechanical devices. In each case, the key step is to identify the relevant high-energy states and their influence on low-energy dynamics through virtual processes.
Finally, we observe that further study is required to increase the accuracy of matrix elements of observables of the matter component.



\vspace{0.5cm}
{\noindent \Large \textbf{Methods}}

\noindent \textbf{Derivation of the effective Hamiltonian in Cavity QED}

In this section, we first present the derivation of the cavity QED Hamiltonian of the main text, while in the next section we focus on the circuit QED case.

We consider the full Hamiltonian expressed in the dipole gauge in \cref{eq:H_full_D}, as the perturbative expansion of the SW transformation is not suitable in the Coulomb gauge (see Supplementary Section 1). To facilitate the application of this procedure, we divide the full Hamiltonian in dipole gauge into three terms:
\begin{equation}
    \label{eq:app-HD_divided}
    \hat{H}_\mathrm{D} = \hat{H}_{0^\prime} + \hat{H}_\mathrm{int, D}^\mathrm{L} + \hat{H}_\mathrm{int, D}^\mathrm{H} \, ,
\end{equation}
where
\begin{equation}
    \label{eq:app-H0_Schrieffer_Wolff}
    \hat{H}_{0^\prime} = \sum_j \hbar \omega^\prime_j \dyad{j}{j}+ \hbar \omega_c \adop \aop
\end{equation}
is the quasi-free Hamiltonian, which includes the high-energy frequencies renormalized by the $\hat{x}^2$ term, i.e.
\begin{equation}
    \label{eq:app-omega_prime}
    \omega^\prime_j = \begin{cases} 
    \omega_j & \text{if } j \leq 1, \\
    \omega_j + \frac{G_{jj}}{\omega_c} & \text{if } j > 1
    \end{cases} \, .
\end{equation}
The light-matter interaction is divided into the low-energy and high-energy terms given by, respectively
\begin{alignat}{2}
    \hat{H}_\mathrm{int, D}^\mathrm{L} =& - i \hbar \sum_{j, k \in \mathcal{S}} g_{jk} \dyad{j}{k} \left( \aop - \adop \right) \nonumber \\
    &+ \hbar \sum_{j,k \in \mathcal{S}} \frac{G_{jk}}{\omega_c} \dyad{j}{k} \label{eq: app - H_int_L} \\
    \hat{H}_\mathrm{int, D}^\mathrm{H} =& - i \hbar \sum_{j,k \in \overline{\mathcal{S}}} g_{jk} \dyad{j}{k} \left( \aop - \adop \right) \nonumber \\
    &+ \hbar \sum_{j,k \neq j \in \overline{\mathcal{S}}} \frac{G_{jk}}{\omega_c} \dyad{j}{k} \, , \label{eq: app - H_int_H}
\end{alignat}
where $\mathcal{S} = \{0,1\}^2$ is the low-energy manifold corresponding to the ground and first excited atomic states, and $\overline{\mathcal{S}} = \mathbb{N}^2 \setminus \mathcal{S}$ is its complementary subspace. This implies that $\hat{H}_\mathrm{int, D}^\mathrm{L}$ has elements only in the low-energy subspace, while $\hat{H}_\mathrm{int, D}^\mathrm{H}$ contains elements in the high-energy subspace.

While the low-energy interaction term cannot be treated perturbatively, the high-energy interaction term $\hat{H}_\mathrm{int, D}^\mathrm{H}$ can be regarded as a perturbation on $\hat{H}_{0^\prime}$. Specifically, we assume that the cavity and the high-energy transitions are in the dispersive regime with the cavity, i.e. $\abs{g_{jk} / (\omega^\prime_{jk} - \omega_c)} \ll 1$. Under this condition, we employ the SW transformation to derive the effective influence of the high-energy terms on the low-energy subspace. 

To this end, we introduce the generator of the SW transformation, $\hat{S}$, which, by using the Baker-Campbell-Hausdorff lemma, satisfies
\begin{equation}
    \label{eq:app-SW-rotation}
    e^{-\hat{S}} \hat{H}_\mathrm{D} e^{\hat{S}} = \hat{H}_\mathrm{D} + \comm{\hat{H}_\mathrm{D}}{\hat{S}} + \frac{1}{2} \comm{\comm{\hat{H}_\mathrm{D}}{\hat{S}}}{\hat{S}} + \ldots \, ,
\end{equation}
By truncating the series to the second order in the perturbation, and subsequently projecting into the two-level subspace, we obtain the effective Hamiltonian in the truncated Hilbert space
\begin{align}
    \hat{\mathcal{H}}_\mathrm{D}^\mathrm{eff} = \hat{P}& \left\{ \hat{H}_{0^\prime} + \hat{H}_\mathrm{int, D}^\mathrm{L} + \hat{H}_\mathrm{int, D}^\mathrm{H} \right. \nonumber \\
    &+ \comm{\hat{H}_{0^\prime}}{\hat{S}} + \comm{\hat{H}_\mathrm{int, D}^\mathrm{L}}{\hat{S}} + \comm{\hat{H}_\mathrm{int, D}^\mathrm{H}}{\hat{S}} \nonumber \\
    &+ \left. \frac{1}{2} \comm{\comm{\hat{H}_{0^\prime}}{\hat{S}}}{\hat{S}} \right\} \hat{P} \, ,
\end{align}
with $\hat{P} = \sum_{j = 0}^{1} \dyad{j}{j}$ the projector onto the two-level subspace. We point out that the series is consistently truncated up to the second order in $A_0$, which is the parameter varied in the plots of the main text.

The generator $\hat{S}$ is defined such that $\comm{\hat{H}_{0^\prime}}{\hat{S}} = - \hat{H}_\mathrm{int, D}^\mathrm{H}$, in order to eliminate the high-energy terms from the effective Hamiltonian. This leads to the following expression for the generator
\begin{align}  \label{eq:app-S}
    \hat{S} =& i \sum_{j,k \in \overline{\mathcal{S}}} \frac{g_{jk}}{\Delta_{jk}} \left[ \omega^\prime_{jk} \left( \aop - \adop \right) + \omega_c \left( \aop + \adop \right) \right] \dyad{j}{k} \nonumber \\
    &- \frac{1}{\omega_c} \sum_{j,k \neq j \in \overline{\mathcal{S}}} \frac{G_{jk}}{\omega^\prime_{jk}} \dyad{j}{k} \, ,
\end{align}
and, consequently, the effective Hamiltonian becomes
\begin{align}  \label{eq:app-H_eff}
    \hat{\mathcal{H}}_\mathrm{D}^\mathrm{eff} = \hat{P} & \left\{ \hat{H}_{0^\prime} + \hat{H}_\mathrm{int, D}^\mathrm{L} \right. \nonumber \\
    &+ \left. \frac{1}{2} \comm{\hat{H}_\mathrm{int, D}^\mathrm{H}}{\hat{S}} + \comm{\hat{H}_\mathrm{int, D}^\mathrm{L}}{\hat{S}} \right\} \hat{P} \, .
\end{align}

We notice that $\hat{P} \comm{\hat{H}_\mathrm{int, D}^\mathrm{L}}{\hat{S}} \hat{P} = 0$, because $\hat{H}_\mathrm{int, D}^\mathrm{L}$ is only composed of terms acting on the two-level subspace, and $\hat{S}$ is composed of terms acting only on its complementary subspace, resulting in a first-order zero contribution in the low-energy subspace. Finally, by explicitly expanding the other terms, we get the following effective Hamiltonian, up to an identity term
\begin{align}
    \label{eq:app-H_eff_complete}
    \frac{\hat{\mathcal{H}}_\mathrm{D}^\mathrm{eff}}{\hbar} =& \omega_c \adop \aop + \left( \frac{\omega_{10}}{2} + \frac{G_{11} - G_{00}}{2 \omega_c} + A_- \right) \sz \nonumber \\
    & + \left( B_+ + B_- \sz \right) \left( \aop - \adop \right)^2 \nonumber \\
    & - i \! \left( g_{01} \! + \! C_{01} \right) \sx \left( \aop - \adop \right) - D_{01} \sy \left( \aop + \adop \right) \, ,
\end{align}
where $A_\pm = A_{11} \pm A_{00}$ and $B_\pm = B_{11} \pm B_{00}$, with the following coefficients:
\begin{alignat}{4}
    \label{eq:app-A_coeff}
    A_{jk} =& \sum_{l>1}^{\infty} \left[ \frac{g_{jl} g_{lk}}{2} \frac{\omega_c}{\Delta_{lk}} - \frac{G_{jl} G_{lk}}{4 \omega_c^2} \left( \frac{1}{\omega^\prime_{lk}} + \frac{1}{\omega^\prime_{lj}} \right) \right] \\
    \label{eq:app-B_coeff}
    B_{jk} =& \sum_{l>1}^{\infty} \frac{g_{jl} g_{lk}}{4} \left( \frac{\omega^\prime_{lk}}{\Delta_{lk}} + \frac{\omega^\prime_{lj}}{\Delta_{lj}} \right) \\
    \label{eq:app-C_coeff}
    C_{jk} =& \sum_{l>1}^{\infty} \frac{1}{2 \omega_c} \left( g_{jl} \frac{G_{lk}}{\omega^\prime_{kl}} + g_{lk} \frac{G_{jl}}{\omega^\prime_{jl}} \right. \nonumber \\
    & \quad \quad + \left. \frac{g_{lk}}{\Delta_{lk}} G_{jl} \omega^\prime_{kl} + \frac{g_{jl}}{\Delta_{jl}} G_{lk} \omega^\prime_{jl} \right) \\
    \label{eq:app-D_coeff}
    D_{jk} =& \sum_{l>1}^{\infty} \frac{1}{2} \left( \frac{g_{jl}}{\Delta_{jl}} G_{lk} - \frac{g_{lk}}{\Delta_{lk}} G_{jl}  \right) \, .
\end{alignat}

The effective Hamiltonian \cref{eq:app-H_eff_complete}, which coincides with \cref{eq:H_QRM_renorm} upon the introduction of the renormalized parameters, describes the low-energy physics of the system in the dipole gauge. However, many of these terms become negligible for increasing anharmonicity. Specifically, we notice that $D_{jk}$ scales as $g_{jl} g_{kl}^2 / \omega_{jl}^2$, where $j$ and $k$ are states in the low-energy subspace and $l$ belongs to the high-energy subspace, while the remaining coefficients scale linearly. Consequently, $D_{jk}$ approaches zero more rapidly compared to the other coefficients. Furthermore, although the individual terms $B_{jj}$ are non-negligible, their difference $B_{11} - B_{00}$ can be neglected as it also scales quadratically with the nonlinearity. With these considerations, we finally get the reduced effective Hamiltonian \cref{eq:H_QRM_D_renormalized_simplified}.

\vspace{0.5cm}
\noindent \textbf{Derivation of the effective Hamiltonian in Circuit QED}

We first rewrite the Hamiltonian in \cref{eq:H_circuit} by using the bosonic operator $\hat{a}$ for the bare mode of the harmonic resonator in \cref{eq:H_res}, while employing the unperturbed states of the fluxonium Hamiltonian in \eqref{eq:H_flux}.
Therefore, the resulting Hamiltonian reads
\begin{align}
    \label{eq:app-H_fluxres}
    \hat{H}_{\rm fr} & = \hbar \omega_c \adop \aop + \hbar \sum_j \omega_j \dyad{j}{j} \nonumber \\
    & - \frac{\phi_\mathrm{zpf}}{L_2} \sum_{j,k} \phi_{jk} \dyad{j}{k} \left( \aop + \adop \right) + \frac{1}{2 L_2} \sum_{j,k} \Phi_{jk} \dyad{j}{k} \, .
\end{align}
By the introduction of the coupling strengths $g_{jk}$ and $G_{jk}$ as in \cref{eq:coupling_circuit} of the main text, the Hamiltonian in \cref{eq:app-H_fluxres} reads
\begin{align}
    \label{eq:app-H_fluxres_expanded}
    \hat{H}_{\rm fr} & = \hbar \omega_c \adop \aop + \hbar \sum_j \omega_j \dyad{j}{j} \nonumber \\
    & - \hbar \sum_{j,k} g_{jk} \dyad{j}{k} \left( \aop + \adop \right) + \hbar \sum_{j,k} \frac{G_{jk}}{\omega_c} \dyad{j}{k} \, .
\end{align}

Both the generator $\hat S$ of the SW transformation and the effective Hamiltonian can be determined by using a key observation: Hamiltonian in \cref{eq:app-H_fluxres_expanded} is structurally equivalent to the dipole gauge Hamiltonian of the cavity QED case in \cref{eq:app-HD_divided}, following the application of the unitary transformation $\hat T = \exp \left( i \pi \adop \aop / 2 \right)$ as $\hat{T} \hat{H}_\mathrm{D}\hat{T}^\dagger$. This transformation interchanges the roles of the position and momentum operators of the resonator, as the only formal difference between the two Hamiltonians lies in the nature of the coupling, as already mentioned in the main text.
Therefore, upon the application of the unitary transformation $\hat T$ to \cref{eq:app-S}, the SW generator for the Hamiltonian of a fluxonium-resonator circuit is
\begin{align} 
    \hat{S} =& \sum_{j,k \in \overline{\mathcal{S}}} \frac{g_{jk}}{\Delta_{jk}} \left[ \omega^\prime_{jk} \left( \aop + \adop \right) + \omega_c \left( \aop - \adop \right) \right] \dyad{j}{k} \nonumber \\
    &- \frac{1}{\omega_c} \sum_{j,k \neq j \in \overline{\mathcal{S}}} \frac{G_{jk}}{\omega^\prime_{jk}} \dyad{j}{k} \, .
\end{align}
Hence, following the same procedure as in the cavity QED case presented in the previous section, the renormalized Hamiltonian is given by

\begin{align} \label{eq:app-H_fluxqubit}
    \frac{\hat{H}_{\rm fq}}{\hbar} & = \omega_c \adop \aop + \left( \frac{\omega_{10}}{2} + \frac{G_{11} - G_{00}}{2 \omega_c} + A_- \right) \sz \nonumber \\ 
    & + \left( \frac{G_{01}}{\omega_c} + A_{10} + A_{01} \right) \sx \nonumber \\
    & - \hbar \left( \frac{\tilde{g}_{11} + \tilde{g}_{00}}{2} \hat{I} + \frac{\tilde{g}_{11} - \tilde{g}_{00}}{2} \sz + \tilde{g}_{01} \sx \right) \left( \aop + \adop \right) \nonumber \\
    & - \hbar \left( B_+ \hat{I} + B_- \sz + 2 B_{01} \sx \right) \left( \aop + \adop \right)^2 \nonumber \\
    & + i \hbar \left(A_{10} - A_{01}\right) \sy \left( \aop^2 - \hat{a}^{\dagger^2} \right) - i \hbar D_{01} \sy \left( \aop - \adop \right) \, ,
\end{align}
where $\tilde{g}_{jk} = g_{jk} + C_{jk}$ and the coefficients are identical to those of natural atoms, as no \emph{a priori} selection rules have been employed in the derivation of the coefficients. However, in evaluating the Hamiltonian in \cref{eq:app-H_eff_complete}, it has been taken into account that the parity of the potential leads to the vanishing of the coupling constants $g_{jk}$ between states of the same parity. Consequently, $G_{jk}$ vanishes between states of different parity. On the other hand, in the circuit QED case, where such symmetry can be broken, several additional terms appear in the renormalized Hamiltonian in \cref{eq:app-H_fluxqubit}.

\vspace{0.5cm}
\noindent {\textbf{Simulation of the $\pi$-pulse}}

To simulate the $\pi$-pulse time evolution in \cref{fig:circuits}(\textbf{f}), the wavefunction of the full system evolves according to the Schr\"odinger equation 
\begin{equation}
    i \hbar \partial_t \ket{\psi} = \hat{H} (t) \ket{\psi} \, ,
\end{equation}
with $\hat{H} (t) = \hat{H}_\mathrm{fr} + \hat{H}_\mathrm{dr} (t)$ and
\begin{equation}
    \hat{H}_\mathrm{dr} (t) = \frac{\hbar \pi}{\sigma_\mathrm{dr} \sqrt{2\pi}} \ e^{- (t - t_0)^2 / 2 \sigma_\mathrm{dr} } \cos (\omega_\mathrm{dr} t) \ \hat{\phi}_1 \, .
\end{equation}
Analogously, in the case of the QRM and RQRM, the time evolution is generated by $\hat{\mathcal{H}}_{\rm fr} + \hat{\mathcal{H}}_{\rm dr} (t)$ and $\hat{\mathcal{H}}_{\rm fr}^{\rm eff} + \hat{\mathcal{H}}_{\rm dr} (t)$, respectively, with
\begin{equation}
    \hat{\mathcal{H}}_{\rm dr} = \hat{P} \hat{H}_\mathrm{dr} \hat{P} =  \frac{\hbar \pi}{\sigma_\mathrm{dr} \sqrt{2\pi}} \ e^{- (t - t_0)^2 / 2 \sigma_\mathrm{dr} } \cos (\omega_\mathrm{dr} t) \ \hat{\sigma}_x \, .
\end{equation}

\vspace{0.5cm}
\noindent {\large \textbf{Acknowledgements}}

We acknowledge enlightening discussions with Filippo Ferrari.
S.S. acknowledges the Army Research Office (ARO) (Grant No. W911NF-19-1-0065).
V.S. acknowledges the Swiss National Science Foundation through
Projects No. 200020\_185015, 200020\_215172, and 20QU-1\_215928.

\vspace{0.5cm}
\noindent {\large \textbf{Code availability}}

The source code generated during the current study is available from the corresponding author upon reasonable request.

\bibliography{bibliography}

\newpage

\setcounter{equation}{0}
\setcounter{figure}{0}
\setcounter{table}{0}
\setcounter{page}{1}
\makeatletter
\renewcommand{\theequation}{S\arabic{equation}}
\renewcommand{\thefigure}{S\arabic{figure}}
\renewcommand{\thepage}{S\arabic{page}}

\onecolumngrid

\begin{center}
\textbf{\Large Supplemental Material for:\\ 
``Renormalization and Low-Energy Effective Models in Cavity and Circuit QED''}
\bigskip
\bigskip
\bigskip
\end{center}

\section{Gauge choice for the Cavity-QED QRM}

The example chosen for the investigation of the renormalization in cavity QED is a single electric dipole of charge $q$ and mass $m$ in a one-dimensional potential, which interacts with a single cavity mode with frequency $\omega_c$.
In the electric-dipole approximation, the radiation wavelength is much larger than the atomic size, allowing us to neglect the spatial dependence of the vector potential, i.e. $\hat A = A_0 \left( \aop + \adop \right)$, where $A_0$ is the zero-point fluctuation amplitude and $\aop \left(\adop\right)$ is the annihilation (creation) operator of the electromagnetic mode. 
The Coulomb gauge Hamiltonian describing such a system reads~\cite{babiker1983derivation,cohen-tannoudji1997photons_lagrangian}
\begin{equation}  \label{eq:sup-H_full_C}
    \hat H_{\rm C} = \frac{( \hat p - q \hat A)^2}{2 m} + V(\hat x) + \hbar \omega_c \adop \aop \, ,
\end{equation}
where $V(\hat x)$ is the atomic potential, $\hat x$ is the position operator and $\hat p$ is its conjugate momentum. Notice the usual minimal coupling replacement $\hat{p} \to \hat{p} - q \hat{A}$ on the atomic momentum, typical of the Coulomb gauge.

Due to the freedom in the choice of the gauge, we can also describe the system using another commonly adopted framework, i.e. the dipole gauge~\cite{power1959coulomb,woolley1971molecular,babiker1983derivation,cohen-tannoudji1997photons_lagrangian}. The two gauges are related by a unitary transformation, given that they represent the same physical system, namely $\hat H_{\rm D} = \hat{U}^\dagger \hat H_{\rm C} \hat U$. In the case of the dipole approximation, $\hat U = \exp ( i q \hat x \hat A / \hbar )$, and the corresponding Hamiltonian reads (Eq.~(1) of the main text)
\begin{align}
    \label{eq:sup-H_full_D}
    \hat H_{\rm D} = & \ \frac{\hat p^2}{2 m} + V(\hat x) + \hbar \omega_c \left( \adop - i \frac{q A_0}{\hbar} \hat{x} \right) \left( \aop + i \frac{q A_0}{\hbar} \hat{x} \right)\nonumber \\
    =& \ \frac{\hat p^2}{2 m} + V(\hat x) + \hbar \omega_c \adop \aop - i q \omega_c A_0 \hat x \omega_c \left( \aop - \adop \right) + \frac{\omega_c q^2 A_0^2}{\hbar} \hat x^2 \, .
\end{align}

The standard QRM is obtained by truncating the atomic Hilbert space to the two lowest energy levels, which can be formally obtained by applying the projection operator $\hat P = \sum_{n=0}^1 \dyad{n}{n}$ to the full Hamiltonian, either in the Coulomb or dipole gauge. 
However, it is well established that these truncated models yield different energy levels, depending on the initial gauge in which the truncation is performed, with the dipole gauge providing a more accurate description within the truncated Hilbert space  \cite{debernardis2018breakdown}.
This discrepancy arises because the interaction term in the Coulomb gauge [see \cref{eq:sup-H_full_C}] depends on the momentum operator $\hat{p}$, whereas in the dipole gauge [see Eq.\eqref{eq:sup-H_full_D}], it depends on the position operator $\hat{x}$.
Since the matrix elements of $\hat{p}$ grow significantly faster than those of $\hat{x}$ increasing the system’s nonlinearity, as given by the relation $\mel{j}{\hat{p}}{k} = i m \omega_{jk} \mel{j}{\hat{x}}{k}$, the Coulomb gauge is less suitable to perform accurate truncation in the non-perturbative regime $g \gtrsim 0.1$ \cite{debernardis2018breakdown}.
This analysis is consistent with Refs.~\cite{distefano2019resolution,savasta2021gauge}, which attributed the poorer accuracy in the Coulomb gauge to the non-locality of the truncated potential. However, the same studies demonstrated how to provide a valid description of the QRM within this gauge, as discussed in the main text.
Therefore, based on the reasons outlined above, we will focus on studying and renormalizing the QRM within the dipole gauge.

\begin{figure*}[t]
    \centering
    \includegraphics{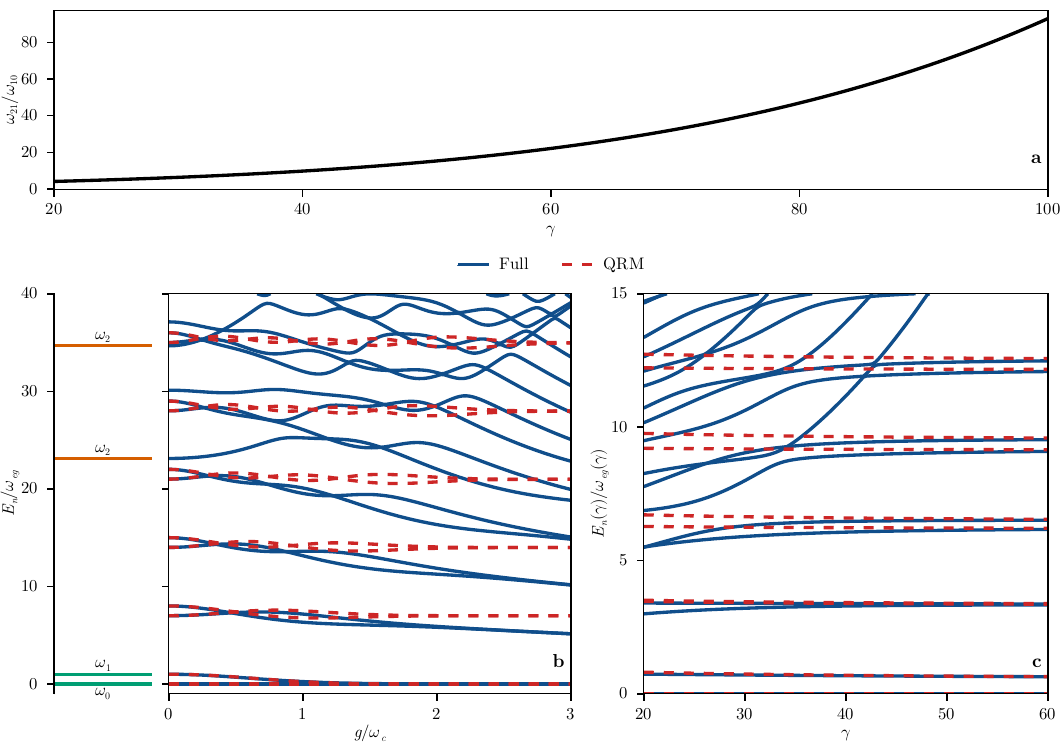}
    \caption{\textbf{Anharmonicity.} \textbf{a} Plot of the frequency ratio $\omega_{21}/\omega_{10}$ as a function of the anharmonicity $\gamma$, highlighting its monotonic increase. \textbf{b} Energy levels comparison between the full model in \cref{eq:sup-H_full_D} and the QRM in \cref{eq:sup-H_QRM}, as functions of the normalized coupling strength $g/\omega_c$. As shown, when the anharmonicity is insufficient compared to the detuning, the third atomic level (left panel) significantly affects the spectral properties, making the two level approximation meaningless. \textbf{c} Energy levels comparison as a function of the anharmonicity, showing progressive convergence as the anharmonicity increases. The parameters used are: $m = 1$, $\gamma = 60$, and $\omega_c = 7 \omega_{10}$ (\textbf{b}); $m = 1$, $g = 1.5 \, \omega_{10}$, and $\omega_c = 3 \omega_{10}$ (\textbf{c}).}
    \label{fig:anharmonicity}
\end{figure*}

The standard QRM Hamiltonian in the dipole gauge reads (Eq.~(2) of the main text)
\begin{equation}  \label{eq:sup-H_QRM}
    \hat{\mathcal{H}}_{\rm D} = \frac{\hbar \bar{\omega}_{10}}{2} \sz + \hbar \omega_c \adop \aop - i \hbar g_{01} \sx \left( \aop - \adop \right) \, ,
\end{equation}
where $\bar{\omega}_{10}$, $g_{jk}$ and $G_{jk}$ have been defined in the main text.
The implications of using the renormalized atomic frequency $\overline{\omega}_{10}$ in the QRM, obtained by projecting onto the bare atomic states, were recently investigated in Ref.~\cite{arwas2023metrics}. In particular, there are three main strategies for projecting the full interacting Hamiltonian: i) Projecting onto the unperturbed atomic eigenstates~\cite{arwas2023metrics}; ii) Projecting onto the eigenstates derived from the effective potential $V_\mathrm{eff} = V(\hat{x}) + \omega_c q^2 A_0^2 \hat{x}^2 / \hbar$~\cite{debernardis2018breakdown}, which incorporates the $\hat{x}^2$ contribution; iii) Projecting directly when implementing the minimal coupling replacement~\cite{distefano2019resolution,savasta2021gauge}. As shown in \cref{eq:sup-H_QRM}, the first approach yields a renormalized atomic frequency $\bar{\omega}_{10}$ originating from the projection of $\hat{x}^2$. The second method still produces a renormalized two-level frequency, although it has a worse accuracy for large detuning. The last approach does not give additional renormalization on the atomic frequency. In this work, we follow the first approach, and the frequency $\overline{\omega}_{10}$ can be interpreted as a preliminary form of renormalization, as it inherently accounts for the contributions of higher-energy states.

\section{Anharmonicity}

In this section, we examine the impact of the anharmonicity on the system's spectrum, using the cavity QED case presented in the main text as a representative example.

As previously discussed, increasing the anharmonicity parameter $\gamma$ leads to an increase in the frequency ratios $\omega_{jk} / \omega_{10}$. Of particular interest are the ratios involving transitions from the two lowest-energy levels to higher excited states, as these mostly determine the validity of the dispersive regime.
For illustrative purposes, \cref{fig:anharmonicity}(\textbf{a}) displays the growth of the ratio $\omega_{21} / \omega_{10}$ with increasing $\gamma$.
\cref{fig:anharmonicity}(\textbf{b}) illustrates how the two-level approximation breaks down when the system's anharmonicity is small compared to the detuning between the cavity and the two-level transition frequency, $\omega_c - \omega_{10}$. In this regime, the third atomic level (shown in the left panel) starts to significantly influence the hybridized energy levels, as evidenced by the growing discrepancies between the full model and the standard QRM, even at low excitation energies.
\cref{fig:anharmonicity}(\textbf{c}) confirms the gradual improvement in the validity of the two-level approximation as the anharmonicity increases. As expected, the lowest-energy levels converge more rapidly, while higher-energy levels require larger values of $\gamma$ to achieve a comparable level of accuracy.

\section{Derivation of the Hamiltonian in Circuit QED}

In this section we focus on the derivation of the full Hamiltonian for the circuit in Fig.~2, which is derived by the use of the standard quantization formalism for circuits. Subsequently, we apply the same procedure outlined in the previous section to derive an effective Hamiltonian for the corresponding two-level approximation.

Taking the two node fluxes $\phi_1$ and $\phi_2$ as the generalized coordinates, the lagrangian of the circuit can be written as
\begin{equation}
    \mathcal{L} = \frac{C_1}{2} \Dot{\phi}_1^2 - \frac{\phi_1^2}{2 L_1} + E_J \cos{\left(\frac{\phi_1 - \phi_{\rm ext}}{\phi_0} \right)} + \frac{C_2}{2} \Dot{\phi}_2^2 - \frac{\left( \phi_2 -\phi_1 \right)^2}{2 L_2}
\end{equation}
where the first line can be interpreted as the lagrangian of the bare fluxonium and the second one can be seen as the lagrangian of the $LC$ resonator with a coordinate-coordinate "minimal coupling", as already pointed out in the main text.
The conjugate momenta are the node charges, given by $Q_i = \frac{\partial \mathcal{L}}{\partial \Dot{\phi}_i} = C_i \Dot{\phi}_i$. Therefore, the Hamiltonian is
\begin{equation}
    \label{eq:sup-H_circ}
    H = \frac{Q_1^2}{2 C_1} + \frac{\phi_1^2}{2 L_1} - E_J \cos{\left(\frac{\phi_1 - \phi_{\rm ext}}{\phi_0} \right)} + \frac{Q_2^2}{2 C_2} + \frac{\left( \phi_2 - \phi_1 \right)^2}{2 L_2}
\end{equation}
We can now promote $\phi_i$ and $Q_i$ to operators satisfying the canonical commutation relation, i.e. $\left[ \hat{\phi}_i , \hat{Q}_j \right] = i \hbar \delta_{ij}$.
Furthermore, we express $\hat{\phi}_2$ and $\hat{Q}_2$, which are associated with the harmonic oscillator, in terms of the bosonic operator $\hat{a}$. On the other hand, due to the anharmonicity of the Josephson junction, $\hat{\phi}_1$ and $\hat{Q}_1$ are expanded in the eigenbasis of the bare fluxonium. 
Thus, Hamiltonian in \cref{eq:sup-H_circ} can be rewritten as
\begin{equation}
    \hat{H} = \hbar \omega_r \adop \aop + \hbar \sum_j \omega_j \dyad{j}{j} - \frac{\hbar \phi_\mathrm{zpf}}{L_2} \sum_{i,j} \phi_{ij} \dyad{i}{j} \left( \aop + \adop \right) + \frac{\hbar}{2 L_2} \sum_{i,j} \Phi_{ij} \dyad{i}{j} \, ,
\end{equation}
which coincides with Eq.~9 of the main text.

\end{document}